\documentclass[
reprint,
superscriptaddress,
amsmath,amssymb
]{revtex4-2}

\usepackage{hyperref}

\usepackage{graphicx}
\usepackage{dcolumn}
\usepackage{bm}
\usepackage{times}
\usepackage[normalem]{ulem} 
\usepackage{orcidlink}

\newcommand{\ie}{i.e., ~}
\newcommand{\eg}{e.g., ~}

\usepackage{xcolor}
\usepackage{color}

\begin{document}


\title{Particle-acceleration mechanisms in multispecies relativistic plasmas}

\author{Claudio Meringolo\orcidlink{0000-0001-8694-3058}}
\affiliation{Institut f\"ur Theoretische Physik, Goethe Universit\"at, Frankfurt, Germany}

\author{Mario Imbrogno\orcidlink{0000-0002-7459-5735}}
\affiliation{Dipartimento di Fisica, Universit\`a della Calabria,
  Arcavacata di Rende, 87036, IT}

\author{Alejandro Cruz-Osorio\orcidlink{0000-0002-3945-6342}}
\affiliation{Instituto de Astronom\'{\i}a, Universidad Nacional
  Aut\'onoma de M\'exico, AP 70-264, 04510 Ciudad de M\'exico, MX}

\author{Sergio Servidio\orcidlink{0000-0001-8184-2151}}
\affiliation{Dipartimento di Fisica, Universit\`a della Calabria,
 Arcavacata di Rende, 87036, IT}

\author{Luciano Rezzolla\orcidlink{0000-0002-1330-7103}}
\affiliation{Institut f\"ur Theoretische Physik, Goethe Universit\"at,
  Max-von-Laue-Str. 1, D-60438 Frankfurt am Main, DE}
\affiliation{School of Natural Sciences and Humanities, New Uzbekistan
  University, Tashkent 100007, Uzbekistan}
\affiliation{School of Mathematics, Trinity College, Dublin, Ireland}

\date{\today}
             
\onecolumngrid
\vspace*{\fill}
\twocolumngrid

\begin{abstract}
While collisionless plasmas are ubiquitously present near astrophysical
compact objects, the impact that their composition has on the high-energy
emission is presently unknown. We present the first investigation of
particle-acceleration mechanisms in kinetic, special-relativistic
turbulence, modeling electrons, positrons, and protons with realistic
mass ratios. Under global charge neutrality, we introduce a positron
fraction and cover regimes ranging from an electron-proton plasma over to
pair-dominated plasmas. Using a novel generalized Ohm's law for
multispecies relativistic plasmas, we analyze particle acceleration due
to electric fields in reconnection events that self-consistently emerge
from turbulence. We demonstrate, for the first time, that energization
occurs at reconnection current sheets driven by the divergence of the
relativistic pressure tensor, which locally aligns with the particle
velocity and leads to an efficient energy transfer. The imbalance between
electrons and positrons systematically favors electron acceleration,
highlighting the necessity of realistic multispecies modeling to capture
the nonthermal contributions in accretion flows and relativistic jets
from black holes.
\end{abstract}

\maketitle


\section{Introduction}Turbulence is a ubiquitous
phenomenon in astrophysical plasma dynamics, often invoked to explain the
origin of nonthermal particles in high-energy astrophysical
sources~\cite{balbus1998instability, biskamp2003magnetohydrodynamic,
  Pezzi2024, Meringolo2024,bacchini2024}. This process plays a central
role in magnetically dominated environments, such as pulsar magnetospheres
and winds, relativistic jets from active galactic nuclei, and coronae of
accretion disks~\cite{Chen2018, Lyutikov2019, Megale2025}. Despite
significant advances in recent decades, driven either by theoretical
models~\cite{Chandran2000, Bruentti2007, Lynn2014} or advanced numerical
simulations~\cite{Dalena2014, Isliker2017, Zhang2023}, the origin of
nonthermal particles in turbulent astrophysical plasmas remains a central
and unresolved problem~\cite{Parker1958, Bell1978, Comisso2018,
  Vega2024}.

In the vicinity of compact objects, extreme conditions of curvature and
electromagnetic fields can lead to the formation of a ``pair-plasma'',
whereby the multispecies plasma consists of electrons, protons and
positrons, whose relative abundances can vary significantly depending on
the specific region considered. The large majority of recent studies of
these important aspects has mostly focused on simplified two-species
plasma models, either neglecting the positron or proton contribution, or
adopting reduced mass ratios to ease computational
constraints~\cite{Sironi2011, Kagan2015, Werner2016, Ball2018}.  Yet,
including the presence of positrons is very important, as it affects the
local charge neutrality, modifies the current structures, and alters the
efficiency of energy-dissipation mechanisms (see, \eg
\citep{Imbrogno2025}), thereby influencing the particle acceleration
processes.

We here present the first investigation of particle acceleration in a
fully self-consistent three-species plasma, using high-resolution,
kinetic Particle-In-Cell (PIC) simulations of turbulence. In all our
simulations, we employ a realistic mass-ratio among electrons, protons,
and positrons and focus attention on the so-called ``trans-relativistic
regime'', where the ratio of magnetic energy density to enthalpy density,
or ``magnetization'', is of the order of unity, \ie $\sigma \sim 1$. In
this configuration, protons remain nonrelativistic, while electrons and
positrons are relativistic.

\begin{figure*}[t]
\centering
\includegraphics[width=2.05\columnwidth]{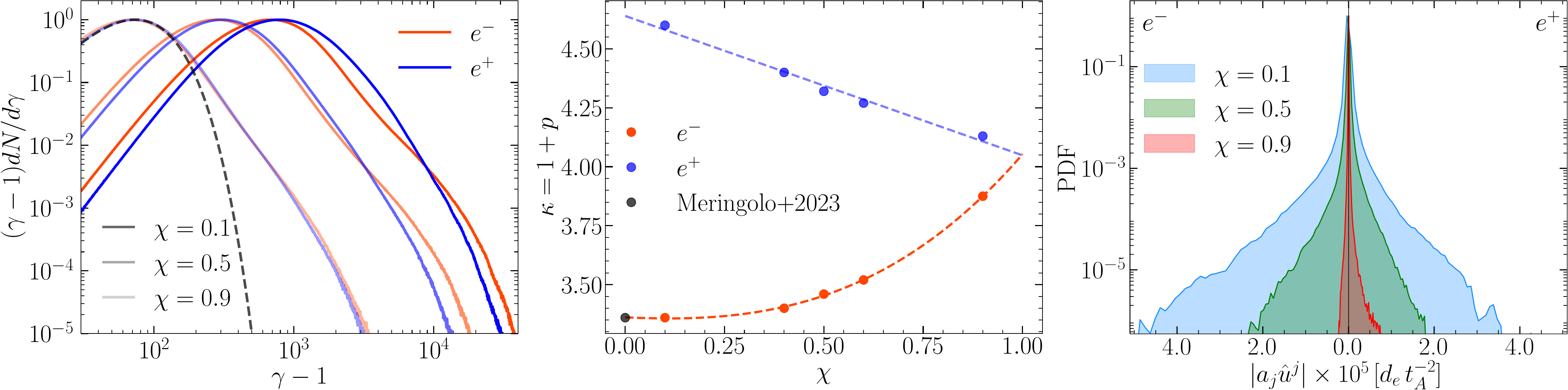}
\caption{\textit{Left:} particle energy spectra at $t/t_A=2.5$ for
  electrons (orange) and positrons (blue), with lines of different
  transparency marking the different mixtures considered ($\chi=0.1,
  0.5$, and $0.9$). The dashed line shows the Maxwell-J\"uttner
  distribution for $\chi=0.9$. \textit{Middle:} variation of the
  power-law index $\kappa$ for electrons (orange) and positrons
  (blue) as a function of plasma mixture $\chi$. The dashed lines
  indicate polynomial fits to the measured behavior. \textit{Right:}
  probability density functions (PDF) of the three-acceleration projected
  along the four-velocity $|a_j \hat{u}^j|$, for electrons (left portion)
  and positrons (right portion), when measured for different values of
  the plasma mixture.}
\label{fig:epsart}
\end{figure*}

\section{Numerical setup}To model different physical
conditions, we introduce the positron fraction parameter $\chi :=
n_{e^+}/n_{e^-}$, defined as the ratio of positron to electron number
densities, with $0 \leq \chi \leq 1$, going from a classical
electron-proton plasma, \ie $\chi \to 0$, to a nearly pure pair plasma,
\ie $\chi \to 1$~\citep{Imbrogno2025}. In all of the simulations, we
assume global charge neutrality, such that the electron number density
balances the combined contribution of protons and positrons, \ie $n_{e^-}
= n_{e^+} + n_{p}$. Furthermore, we employ a two-dimensional (2D)
geometry while retaining the full three-dimensional components of the
electromagnetic fields, as required for any consistent plasma model.  The
plasma is initially uniform and follows a relativistic Maxwell-J\"uttner
distribution with a thermal spread that is characterized by the
dimensionless temperature $\theta_\alpha := T_\alpha/m_\alpha$, where
$\alpha = \{e^-, e^+, p \}$, to refer to electrons, positrons, and
protons, respectively. The turbulence is seeded through uncorrelated
magnetic fluctuations imposed via a perturbed 2D Fourier spectrum of the
magnetic field~\cite{Meringolo2023, Imbrogno2024} (see End Matter). On
the particle side, the initial distribution is characterized by its
thermal Lorentz factor, which is initially proportional to the
magnetisation \ie $\gamma_{\rm th} \simeq \sigma$, and increases with
larger initial temperatures $\theta_\alpha$~\cite{Comisso2019,
  Wong2020}. On the other hand, the maximum Lorentz factor in a turbulent
plasma is set by $\gamma_{\rm max} \simeq \langle \bm{B} \rangle \ell_c
$, where $\ell_c :=\int C(x)dx$ is the coherence lengthscale of the
turbulence and $C(x)$ is the correlation function~\cite{Pecora2018}, so
that for energies for which $\gamma > \gamma_{\rm max}$ particles
essentially decouple from the turbulent motion~\cite{Hillas1984,
  Lemoine2025}.

Once the simulations begin, turbulence rapidly develops, producing a
complex landscape characterized by strong intermittency and large spatial
variability~\cite{Servidio2012}, as compressibility is enhanced by
relativistic effects~\citep{Radice2012b}. In addition, relativistic
turbulence naturally gives rise to extreme particle acceleration events,
yielding sizeable populations of nonthermal particles, particularly
electrons and positrons~\cite{Comisso2018, Lemoine2025}.

\section{Results}The acceleration process of
different particles is illustrated in Fig.~\ref{fig:epsart}, whose left
panel reports the energy distribution functions (EDFs), \hbox{$(\gamma-1)
  dN/d \gamma$} as a function of $\gamma-1$, at time\footnote{Here, $t_A
:= L / v_A$ is the Alfv\'en time, with $v_A := \sqrt{\sigma/(1+\sigma)}$
the Alfv\'en speed.} $t/t_A = 2.5$ for representative values of
$\chi$. Here, \hbox{$\gamma := \sqrt{1+u^2}$} is the Lorentz factor and
$u^j=\gamma v^j$ are the spatial components of the particle four-velocity
with three-velocity $v^j$. Specifically, we show EDFs for electrons
(orange solid lines) and positrons (blue solid lines), with different
shades corresponding to the different mixtures, \ie $\chi = 0.1, 0.5$,
and $0.9$. As a reference, the black dashed line represents the
Maxwell-J\"uttner distribution that best fits the low-energy part of the
EDFs, \ie the thermal population, for the case $\chi=0.9$.

We note that the high-energy tail of the spectra is well approximated by
a power law $dN/d\gamma \propto \gamma^{-p}$~\citep{Davelaar2019,
  Fromm2022}, whose spectral indices $\kappa := 1 + p$ differ between
electrons and protons. This is nicely summarized in the middle panel of
Fig.~\ref{fig:epsart}, which reports the changes in $\kappa =
\kappa(\chi)$ for different mixtures of $e^+$ (blue solid circles) and of
$e^-$ (orange solid circles), and where the dashed lines show polynomial
fits to the data given by $\kappa_{e^-} (\chi)= 0.48\chi^3 + 0.29\chi^2 -
0.08\chi + 3.36$ and $\kappa_{e^+} (\chi)= -0.59 \chi + 4.63$. In the
extreme case of a pure electron-proton plasma, \ie $\chi \rightarrow 0$,
our results are in agreement with those reported in
Ref.~\cite{Meringolo2023} (black point in middle panel), while
for a pure pair plasma, \ie $\chi \rightarrow 1$, the two spectra converge to
the same value, $\kappa_{e^+}(\chi \rightarrow 1) \simeq \kappa_{e^-}(\chi
\rightarrow 1) \simeq 4.005$. While the convergence obviously reflects
the symmetry in the mixture, the specific value for $\kappa$ is instead
the result of the specific choice made for $\beta_{e^-}$ and
$\sigma$. The right panel shows instead the probability density function
(PDF) of the three-acceleration projected along the four-velocity, $|a_j
\hat{u}^j| = |a_j v^j/\sqrt{v_i v^i}|$, for electrons (left part) and
positrons (right part), computed over the duration of each
simulation. Different shaded regions correspond to different values of
$\chi$, revealing that electrons systematically reach higher
accelerations than positrons. This trend is consistent with the EDFs
shown in the left panel, where electrons exhibit a global shift toward
higher $\gamma$ as $\chi$ decreases, along with a more pronounced
nonthermal tail at higher energies compared to positrons. These
differences in acceleration behavior can be understood in terms of the
local electric field configuration, as we discuss below.

By taking velocity moments of the multispecies collisionless plasma, one
can derive a generalized Ohm's law, where the total electric field
separates into distinct large- and small-scale contributions as $\bm{E}
\simeq \bm{E}_{\bm{V} \times \bm{B}} + \bm{E}_{\bm{\nabla} \cdot
  \bm{\Pi}}$, where~\cite{Imbrogno2025}
\begin{align}
  \bm{E}_{\bm{V} \times \bm{B}} := & -\frac{1}{\mathcal{N}} \big[ n_{e^+} \bm{V}_{e^+}
  + n_{e^-} \bm{V}_{e^-} \big] \times \bm{B}\,, 
  \label{E1}  \\
  \bm{E}_{\bm{\nabla} \cdot \bm{\Pi}} := & \frac{1}{e \, \mathcal{N}} \,
  \big[ \bm{\nabla} \cdot \bm{{\bm{\Pi}}}_{e^+} - \bm{\nabla} \cdot
    \bm{{\bm{\Pi}}}_{e^-} \big]\,.
  \label{E2}
\end{align}
Here, $\mathcal{N} := n_{e^+} + n_{e^-}$ denotes the total number density
of positrons and electrons, while $\bm{V}_\alpha$ represents the bulk
three-velocity of species $\alpha$. The generalized pressure tensor is
defined as $\bm{{\bm{\Pi}}}_{a} := \int f_{a} m_{a} \bm{u} \bm{u} /
\gamma \, d^3u$, where $f_{a} = f_{a} (\bm{x}, \bm{u}, t)$ is the
particle distribution function of the $\alpha$-th species, with position
$\bm{x}$ and four-velocity $\bm{u}$. We note that the contribution from
the slowly moving protons has been neglected in the convective electric
field given by Eq.~(\ref{E1}), while the small-scale electric field in
Eq.~(\ref{E2}), typically associated with non-ideal effects in collisional
plasmas, can also develop, in principle, a component parallel to the
local magnetic field.

\begin{figure}
\includegraphics[width=0.9\columnwidth]{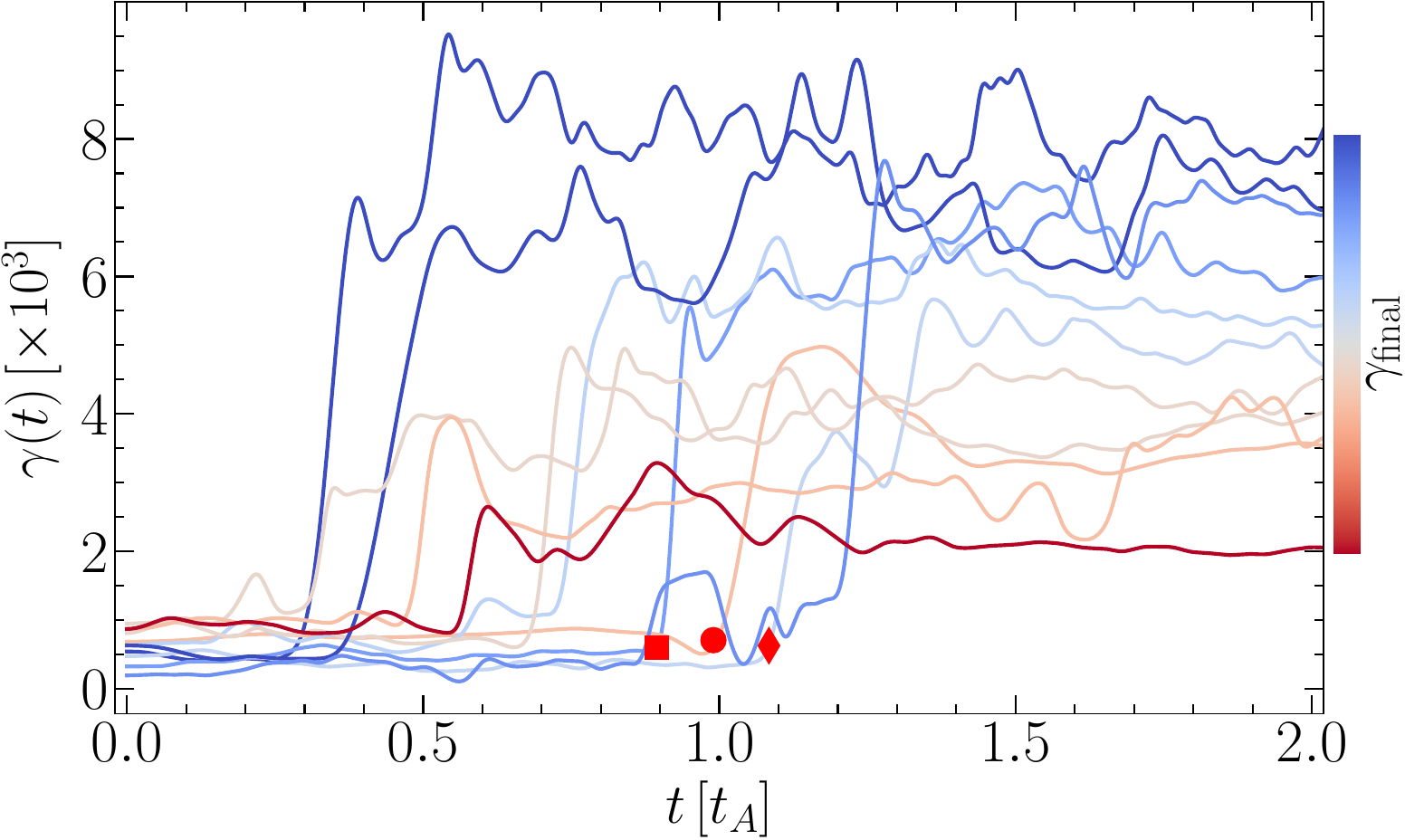}
\vskip 0.3cm
\includegraphics[width=0.9\columnwidth]{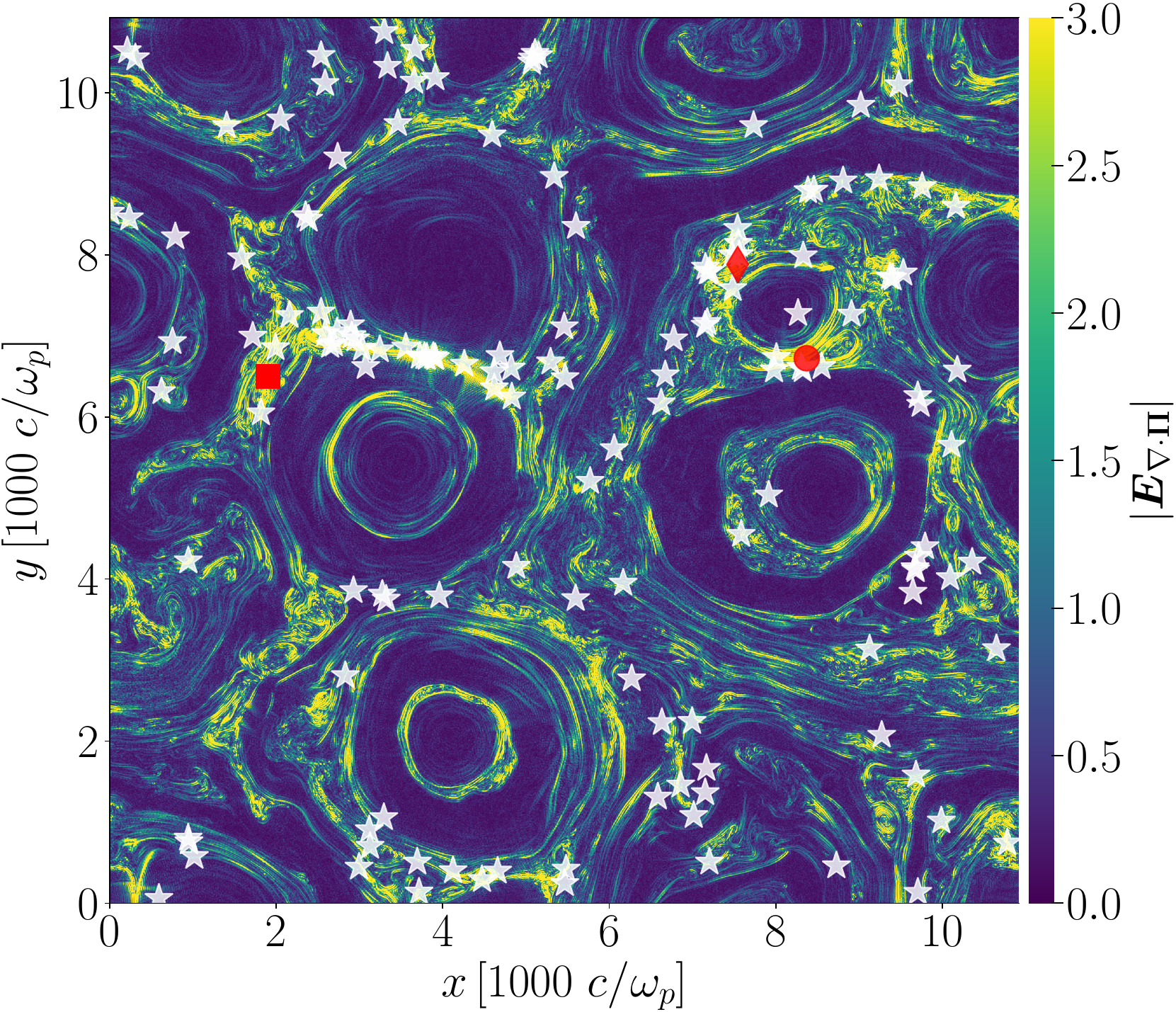}
\caption{\textit{Top:} evolution of the Lorentz factor for a set of
  representative particles that receive a substantial acceleration at one
  point in time $t_*$; the red symbols refer to particles whose
  spatial position is marked below. The data refers to the mixture with
  $\chi = 0.5$ and the final Lorentz factor are expressed with a
  colormap. \textit{Bottom:} 2D map of the magnitude of the electric
  field $\bm{E}_{\bm{\nabla} \cdot \bm{\Pi}}$ at time $t/t_A =
  1.0$. White stars indicate the positions of the most energetic
  particles at the time of the acceleration $t_*$ within the time
  interval $0.9 \leq t/t_A \leq 1.1$. Red symbols mark the spatial
  location of the electrons reported in the top panel.}
\label{fig:2}
\end{figure}

To unveil the mechanisms behind particle acceleration, we tracked the
trajectories of a random sample of $\sim 10^4$ particles. The top part of
Fig.~\ref{fig:2} shows the evolution of the Lorentz factor $\gamma(t)$
for a few representative particles in the case $\chi = 0.5$, which
eventually populate the nonthermal tail of the energy distribution (see
the left panel of Fig.~\ref{fig:epsart} for $\gamma \lesssim 10^4$). A
common feature among these particles is a sudden and rapid energy gain,
characterized by an abrupt, impulsive increase in the Lorentz
factor. This explosive acceleration is highlighted at time $t \sim t_*$
(red symbols in Fig.~\ref{fig:2} for three representative particles),
where $t_*$ marks the onset of the primary acceleration episode, defined
as the instant when the particle experiences an increase in the Lorentz
factor $d\gamma/dt > 10^4 \, t_A^{-1}$. We further isolated the most
energetic particles that satisfy the above condition within the time
window $0.9 \leq t/t_A \leq 1.1$. Their positions at $t = t_*$ are marked
with white stars in Fig.~\ref{fig:2} (bottom panel), while the red
symbols refer to the particles shown in the top panel.  The colormap
displays the magnitude of the contribution of the electric field
$\bm{E}_{\bm{\nabla} \cdot \bm{\Pi}}$ as defined in Eq.~(\ref{E2}).

In these regions, strong velocity shears generate large gradients in the
pressure tensor, leading to a systematic imbalance among species. This
imbalance originates from the global charge neutrality condition in a
three-species plasma. Stated differently, the presence of protons (which
behave as an essentially unmagnetized, slow-moving fluid) determines a
reduction in positron density. Consequently, the positron-pressure
contribution is much smaller than the corresponding electron-pressure
contribution. This pressure asymmetry enhances local non-ideal electric
fields, resulting in more efficient particle acceleration and higher
reconnection rates as $\chi \rightarrow 1$. Conversely, near the centers
of magnetic islands, where the magnetic field lines are topologically
closed, the plasma approaches isotropy and the particle distribution
tends to a thermal one, the electron and positron pressures are
comparable and the electric field $\bm{E}_{\bm{\nabla} \cdot \bm{\Pi}}$
becomes negligible.

Particle acceleration in regions of active reconnection generally arises
from the interaction of the particle orbits with small-scale current
sheets~\cite{Drake2009}. This mechanism extracts particles from the
thermal population and accelerates them to high energies through primary
and secondary Fermi-like processes~\citep{Zank2015, Pecora2018}. Under
these conditions, acceleration is initially driven by electric fields
aligned with the magnetic field, $\bm{E}_{\parallel}$, and is later
enhanced by the perpendicular component $\bm{E}_{\perp}$ through
stochastic scattering off turbulent fluctuations~\cite{Comisso2018}.

\begin{figure}[t]
\includegraphics[width=0.9\columnwidth,height=0.9\columnwidth]{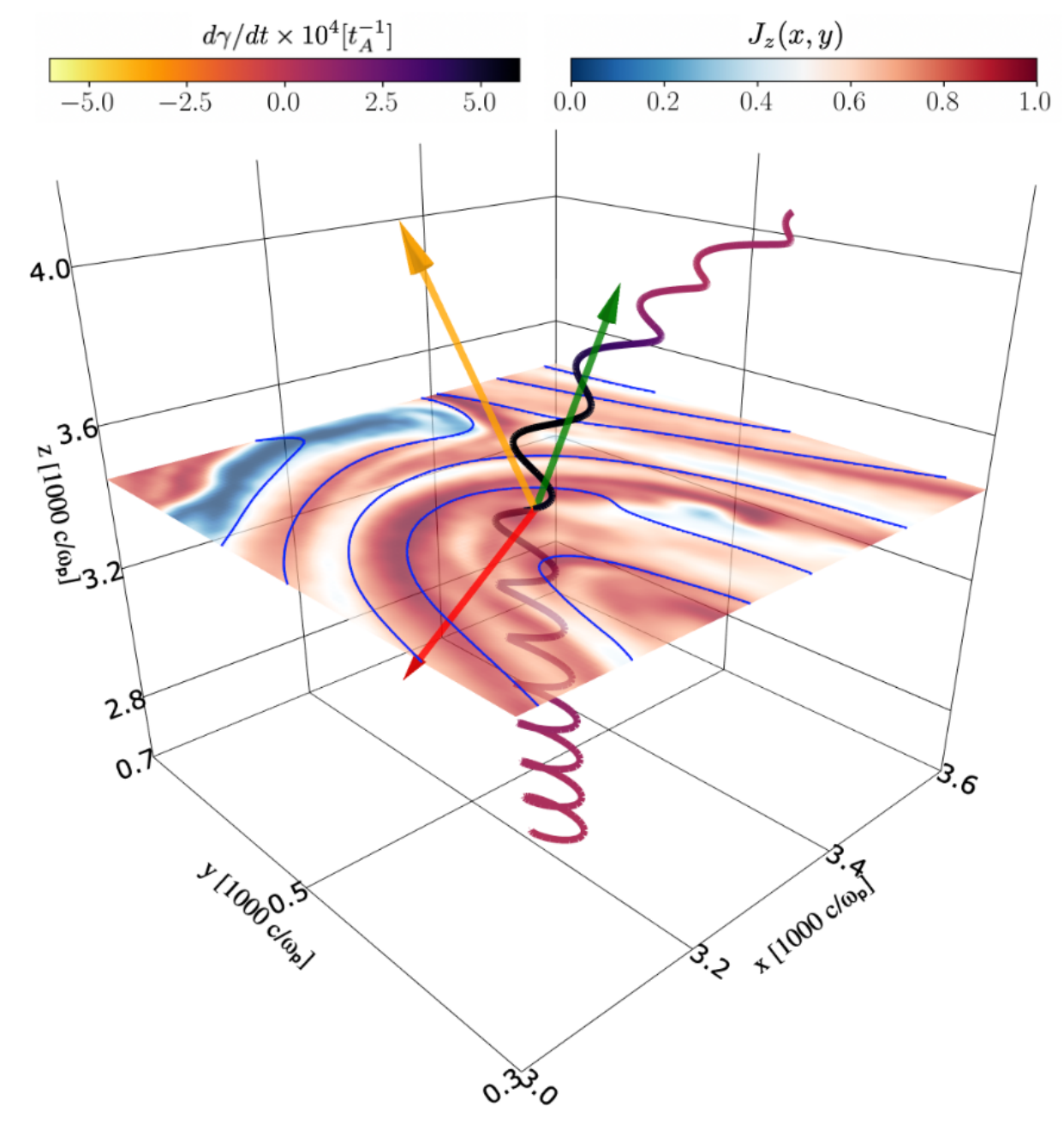}
\caption{3D trajectory of a representative $e^-$ particle accelerated by
  magnetic reconnection, for the case $\chi=0.5$. The colorcode along the
  particle trajectory is related to the change in energy $d \gamma / dt$,
  while the $z={\rm const.}$ 2D slice shows with a colormap the
  out-of-plane current density $J_z$ at the time $t_* \sim 0.74 \, t_A$,
  when the particle is accelerated.  Blue solid lines on the 2D slice
  mark the iso-contours of the vector potential $A_z$, while the colored
  arrows indicate the local magnetic field $B$ (green) and the two main
  components of the electric field, $\bm{E}_{\bm{\nabla} \cdot \bm{\Pi}}$
  (red) and $\bm{E}_{\bm{V} \times \bm{B}}$ (orange).}
\label{fig:3d_1}
\end{figure}

Figure~\ref{fig:3d_1} concentrates on a representative electron
undergoing sudden acceleration for the intermediate case $\chi =
0.5$. The left panel shows a 3D reconstruction of the particle trajectory
in the time range $0.65 \leq t/t_A \leq 0.85$, color-coded by the rate of
change of the particle energy $d\gamma / dt$. Initially, $d\gamma / dt
\sim 0$, so that the electron experiences only negligible
acceleration. However, around $t_*/t_A \simeq 0.74$ it undergoes a sharp
energization, reaching $d\gamma / dt \sim 6 \times 10^4 \, t_A^{-1}$.
Also reported with a colormap is a 2D slice of the current density $J_z$,
with the blue solid contours tracing iso-contours of the vector potential
$A_z$. In this way, it is possible to appreciate that the acceleration
event occurs within a current sheet, that is, an active reconnection site
where particles are accelerated to nonthermal
energies~\cite{Bacchini2019, Sironi2014}. Also shown with arrows at the
time of the acceleration are the vector fields referring to the magnetic
field $\bm{B}$ (green) and to the two components of the electric field,
\ie $\bm{E}_{\bm{V} \times \bm{B}}$ (orange) and $\bm{E}_{\bm{\nabla}
  \cdot \bm{\Pi}}$ (red).

\begin{figure}[t]
\includegraphics[width=0.9\columnwidth,height=0.9\columnwidth]{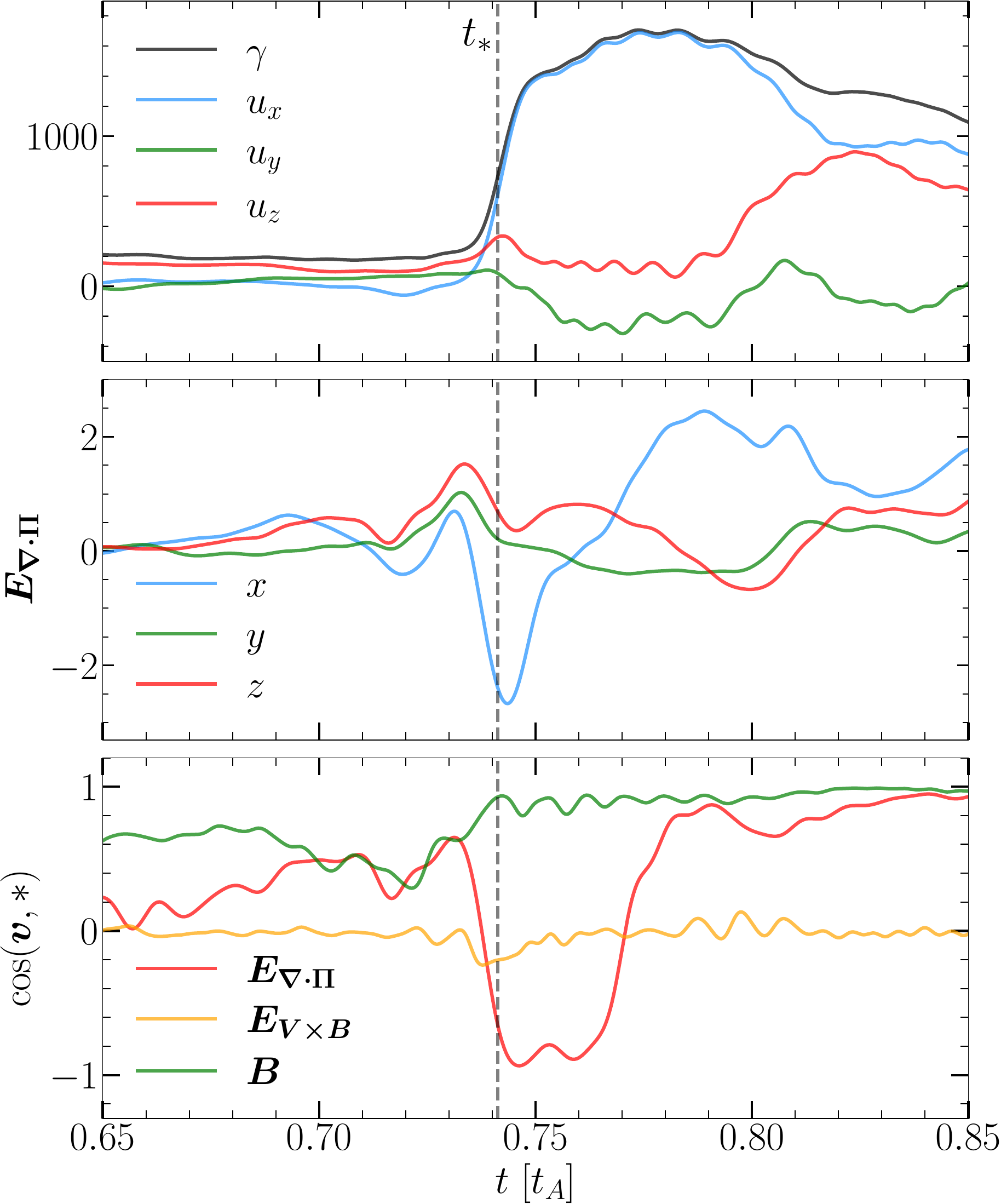}
\caption{Evolution of the particle Lorentz factor and four-velocity
  components (top); components of $\bm{E}_{\bm{\nabla} \cdot \bm{\Pi}}$
  (middle), and the cosine of the angle between the colored vectors on
  the left panel and the particle trajectory (bottom).}
\label{fig:3d_2}
\end{figure}

A more quantitative description is offered in Fig.~\ref{fig:3d_2}, which
describe several properties of the particle shown on the left. In
particular, the top panel reports the evolution of the Lorentz factor and
four-velocity components and shows that the energization event takes
place with the dynamical properties changing mostly in the (in-plane) $x$
direction. Furthermore, the middle panel highlights that at the time of
the strong acceleration, $t_*$, the particle experiences a strong
electric field due to the pressure-tensor divergence,
$\bm{E}_{\bm{\nabla} \cdot \bm{\Pi}}$. Furthermore, to clarify which of
the possible accelerating mechanism is the most relevant one, the bottom
panel reports the cosine of the angle between the particle velocity and
the various accelerating fields (bottom panel). In this way, it is
possible to realize that at $t_*$ the velocity aligns strongly with the
electric field associated with $\bm{E}_{\bm{\nabla} \cdot \bm{\Pi}}$,
while remaining nearly perpendicular to the large-scale convective field
$\bm{E}_{\bm{V} \times \bm{B}}$ and to the magnetic field
$\bm{B}$. Interestingly, while what we discussed so far referred to
electrons, the same applies also when considering the acceleration of
positrons.

\begin{figure*}[t]
\includegraphics[width=2.08\columnwidth]{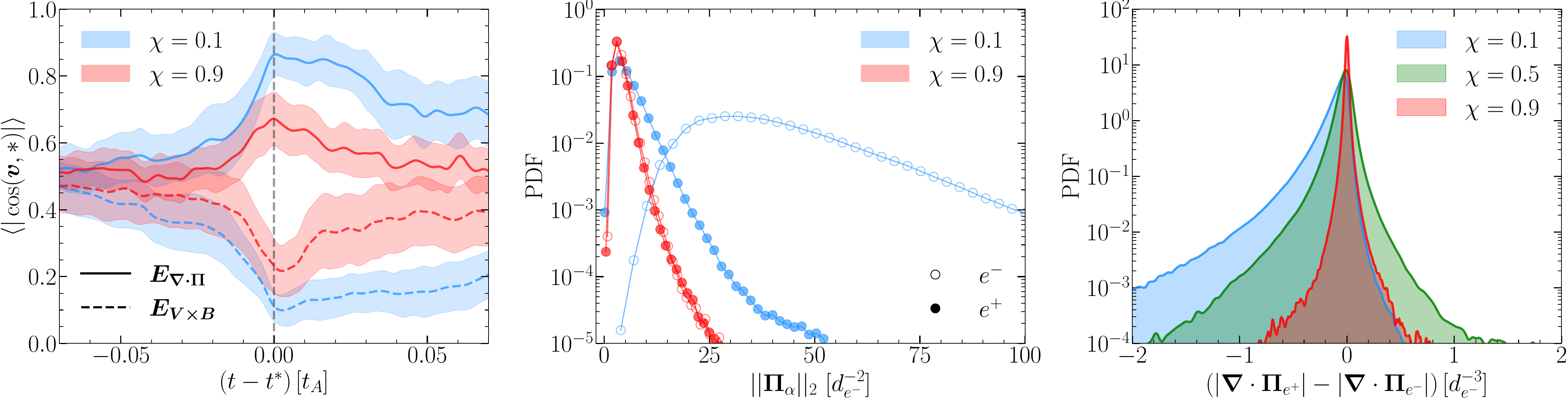}
\caption{\textit{Left:} average alignment between electron trajectories
  and the electric fields given by cosines $|\cos(\bm{v}, \bm{*})|$, with
  $\bm{*}$ being either $\bm{E}_{\bm{\nabla} \cdot \bm{\Pi}}$ (solid
  lines) or $ \bm{E}_{\bm{V} \times \bm{B}}$ (dashed lines). The data
  refers to $\chi=0.1$ (blue colors) and $\chi=0.9$ (red colors), with
  the shaded areas representing the variance in the measurement.
  \textit{Middle:} PDFs of the $L_2$-norm of the generalized pressure
  tensor $\bm{\Pi}_\alpha$ relative to electrons (open circles) and
  positrons (filled circles), with colors denoting different values of
  $\chi$. \textit{Left:} PDFs of the differential acceleration
  quantified by $|\nabla \cdot \bm{\Pi}_{e^+}| - |\nabla \cdot
  \bm{\Pi}_{e^-}|$, where different colors refer to different plasma
  mixtures.}
\label{fig:4}
\end{figure*}

While Figs.~\ref{fig:3d_1} and \ref{fig:3d_2} provide a transparent
representation of what happens to an electron in the case of large
energization process, it is important to assess the robustness and
universality of this acceleration mechanism. To this scope, among all the
$10^4$ particles tracked in \texttt{chi0.1}, \texttt{chi0.5}, and
\texttt{chi0.9} (see Tab.~\ref{tab:table1} in the End Matter), we
analyzed the evolution of those for which $d\gamma/dt > 10^4 \, t_A^{-1}$
at one point in time.  More specifically, for each selected particle, we
measured the alignment between its velocity and the electric-field
components defined in Eqs.~(\ref{E1}) and~(\ref{E2}), focusing on
particles that undergo sudden acceleration at $t=t_*$ and initially have
Lorentz factors $\gamma(t_*) \lesssim \gamma_{\rm th}$. We note that it
is essential for this investigation to store information with a high
cadence, with particle trajectories being recorded at every time step
(\ie $\Delta t \simeq 6.5 \times 10^{-5} \, t_A$) and with fluid
quantities being stored every $\sim 150$ steps, (\ie with $\Delta t / t_A
\simeq 0.01$).

The results of this more selected particle analysis are presented in the
left panel of Fig.~\ref{fig:4}, which reports the average alignment
between electron trajectories and the electric fields, quantified by the
cosines $|\cos(\bm{v}, \bm{*})|$, with $\bm{*}$ being either
$\bm{E}_{\bm{\nabla} \cdot \bm{\Pi}}$ (solid lines) or $ \bm{E}_{\bm{V}
  \times \bm{B}}$ (dashed lines). The data refers to two representative
values of the positron fraction, namely, $\chi=0.1$ (blue colors) and
$\chi=0.9$ (red colors); for each line and color group, the shaded area
represents the variance in the measurement. 

In this way, it is possible to appreciate that particles begin to align
with $\bm{E}_{\bm{\nabla} \cdot \bm{\Pi}}$ a few $0.01\, t_A$ before the
acceleration time $t_*$, reaching peak correlation at the moment of
maximum acceleration. For $t \gtrsim t_*$, the alignment gradually
weakens, with the cosine approaching an asymptotic value of $\sim 0.5$ --
consistent with a random angle distribution after long
transients. Conversely, the $\bm{E}_{\bm{V} \times \bm{B}}$ component
becomes nearly perpendicular to the particle velocity as $t \rightarrow
t_*$, signaling the onset of a stochastic acceleration phase dominated by
scattering off turbulent fluctuations. This behavior is qualitatively and
quantitatively similar for both electrons and positrons (not shown).

The middle panel of Fig.~\ref{fig:4} shows the PDFs of $L_2$-norm of the
generalized pressure tensor $\bm{\Pi}_\alpha$ sampled by the particles at
$t = 2 \, t_A$, for electrons, \ie $\bm{\Pi}_{e^-}$ (open symbols) and
positrons, \ie $\bm{\Pi}_{e^+}$, (solid symbols). Also in this case, the
data refers to two different positron fractions, namely, $\chi=0.1$ (blue
colors) and $\chi=0.9$ (red colors). Note that the low-$\chi$ plasma
contains significantly fewer positrons than electrons, so that from a
fluid perspective, this subdominant species contributes less to the total
pressure, resulting in $\langle \bm{\Pi}_{e^-} \rangle \gg \langle
\bm{\Pi}_{e^+} \rangle$ where $\langle \cdot \rangle$ refers to the median of
the PDF. As a result, at low $\chi$, electrons experience stronger
accelerations and exhibit a PDF which is much broader than that of the
positrons. By contrast, at higher positron fractions, the pressure
contributions nearly coincide (red curves) and the PDFs are very similar,
as expected in a symmetric pair-plasma limit.

The pressure imbalance produces a fundamental difference in the
acceleration mechanism, which can only be attributed to the electric
field parallel to the local magnetic field and associated with the
pressure-tensor divergence. To measure this imbalance, we define a global
measure of the acceleration as $\bm{\mathcal{A}}_{e^\pm} := \int
\bm{a}_{e^\pm} \, f_{e^\pm}\, d^3u = \pm \, n_{e^\pm} (\bm{E} +
\bm{V}_{e^\pm} \times \bm{B})$ where $\bm{a}_{e^\pm} := d \bm{u}_{e^\pm}
/ dt$ is the single-particle acceleration. The difference in acceleration
between species can then be expressed as
\begin{equation}
\bm{\mathcal{A}}_{e^+} \! - \!\bm{\mathcal{A}}_{e^-} \simeq \nabla \cdot
\bm{\Pi}_{e^+}\! -\!  \nabla \cdot \bm{\Pi}_{e^-}\,,
\end{equation}
where we used Eq.~(\ref{E2}) in evaluating the Lorentz force and assumed,
for simplicity, that $\bm{E} \simeq \bm{E}_{\bm{V} \times \bm{B}} +
\bm{E}_{\bm{\nabla} \cdot \bm{\Pi}}$. In the right panel of
Fig.~\ref{fig:4} we report the PDF of $|\nabla \cdot \bm{\Pi}_{e^+}| -
|\nabla \cdot \bm{\Pi}_{e^-}|$, for three different mixtures of
$\chi=0.1, 0.5$, and $0.9$. The distributions exhibit a clear asymmetry,
with negative skewness $\mathcal{S}=-10.4, -3.3$ and $-1.2$,
respectively, that decreases as $\chi \rightarrow 1$, approaching
symmetry in the pair-plasma limit. Overall, this information provides
strong confirmation that symmetry breaking in particle acceleration is
directly controlled by the plasma composition.

\section{Conclusions}We have presented the first
systematic analysis of 2D kinetic turbulence in relativistic plasmas
composed of three species of charged particles: electrons, positrons, and
protons, with a realistic mass ratio. We believe this represents a
significant step forward in modeling astrophysical plasmas near compact
objects, where all three species are expected to coexist.

Our analysis has revealed that the electron energy spectrum depends
sensitively on the positron fraction $\chi$, with a significant
nonthermal component characterized by harder power-law slopes emerging as
$\chi$ decreases, \ie as the mixture tends to that of proton-electron
plasma. Using a generalized form of Ohm's law for multispecies
relativistic plasmas~\cite{Imbrogno2025}, we have also investigated in
detail the mechanisms responsible for particle acceleration operating
during reconnection events. In this way, we have found that particles
initially at thermal energies -- low Lorentz factors -- are accelerated
at reconnection current sheets, with subsequent motions that are aligned
in the direction given by the electric field generated by the divergence
of the pressure tensor. This efficient acceleration and energization
process is the one responsible for the creation of the significant
nonthermal tails. Although the alignment is present for all mixtures, it is
stronger for lower values of $\chi$, with electrons and positrons exhibiting
a broadly similar behavior.

While our findings open the way to a more extensive exploration of the
properties of realistic astrophysical plasmas and provide motivation for
a range of follow-up studies, they will also benefit from a number of
improvements. Besides the obvious use of even higher spatial and particle
resolutions, important advances in our understanding of multispecies
relativistic turbulence will come with the extension to three spatial
dimensions \cite{Nathanail2021b} and the addition of general-relativistic 
effects \citep{Vos2024, Meringolo2025a, Meringolo2025b}. All of these
aspects will be the focus of future work.

\vspace{0.25cm}
\subsection*{Acknowledgements}
This research is supported by the
ERC Advanced Grant ``JETSET: Launching, propagation and emission of
relativistic jets from binary mergers and across mass scales'' (Grant
No. 884631), the European Union's Horizon Europe research and innovation
programme under grant agreement No. 101082633 (ASAP). A.C.O. acknowledges
DGAPA-UNAM (grant IN110522) and the Ciencia B\'asica y de Frontera
2023-2024 program of SECIHTI M\'exico (projects CBF2023-2024-1102 and
257435). L.R. acknowledges the Walter Greiner Gesellschaft zur
F\"orderung der physikalischen Grundlagenforschung e.V. through the Carl
W. Fueck Laureatus Chair. Computational resources were provided by CINECA
through the ISCRA Class B project ``KITCOM-HP10BB7U73''. We acknowledge
ISCRA for awarding this project access to the LEONARDO supercomputer,
owned by the EuroHPC Joint Undertaking, hosted by CINECA
(Italy). Simulations were also performed on HPE Apollo HAWK at the High
Performance Computing Center Stuttgart (HLRS) under the grant BNSMIC, on
the Goethe-HLR supercomputer, and on the local supercomputing cluster
Calea.

\section*{Appendix}

\label{appendix1}

In our kinetic approach, we solve numerically the Vlasov-Maxwell
equations using a modified version of the special-relativistic
\texttt{Zeltron} code~\cite{Cerutti2013}, capable of handling a
three-species plasma. All simulations are initialized in a square
Cartesian domain of side $L_x = L_y =: L \simeq 10923 \, d_{e^-}$, where
$d_{e^-} := \sqrt{m_{e^-}/(4 \pi e^2 n_{e^-})}$ is the electron-skin
depth, $e$ and $m_{e^-}$ are the electron charge and mass,
respectively. The computational grid consists of $N_x = N_y = N = 8192$
mesh points, with periodic boundary conditions, where each cell is
initialized with $240$ particles (or ``macroparticles''), yielding $\sim
10^{10}$ particles per simulation to minimize numerical noise. We adopt
units where $e = m_{e^-} = 1$ and have summarized all key simulation
parameters in Tab.~\ref{tab:table1}, where $\beta_{e^-} = \beta_{e^+} +
\beta_{p}=0.1$ and the magnetization is set to be $\sigma \sim 1$ in all
the cases.

\begin{table}[t]
  \begin{ruledtabular}
    \begin{tabular}{cccccccc}
      mixture & $\chi$ & $\beta_{e^+}$ & $\beta_{p}$ & $\theta_{e^-}, \theta_{e^+}$ & 
      $\theta_{p}$ & $B_0$ & $PP\lambda^2_D$ \\
      \hline
      $\texttt{chi0.1}$& 0.1 & 0.01 & 0.09 & 137.8 & 0.075 & 52.50 & 18585 \\
      $\texttt{chi0.4}$& 0.4 & 0.04 & 0.06 & 91.92 & 0.050 & 42.88 & 12407 \\
      $\texttt{chi0.5}$& 0.5 & 0.05 & 0.05 & 76.63 & 0.042 & 39.15 & 10359 \\
      $\texttt{chi0.6}$& 0.6 & 0.06 & 0.04 & 61.33 & 0.034 & 35.02 & ~8269 \\
      $\texttt{chi0.9}$& 0.9 & 0.09 & 0.01 & 15.46 & 0.008 & 17.58 & ~2088 \\
    \end{tabular}
  \end{ruledtabular}
  \caption{Summary of the simulation properties. From left to right, the
    different columns report: the global positron-to-electron
    concentration ratio $\chi$, the plasma beta parameter for positrons
    $\beta_{e^+}$ and protons $\beta_p$, the dimensionless temperature
    for the electrons $\theta_{e^-}$, the positrons $\theta_{e^+}$, and
    the protons $\theta_{p}$; the out-of-plane magnetic-field strength,
    $B_{0}$, and the total number of particles per unit Debye area
    $PP\lambda_{D}^{2}$. }
  \label{tab:table1}
\end{table}

The fluctuations are initialized by expressing the $z$-component of the
vector potential in terms of Fourier modes as $A_z(x, y) := \sum_{k_x,
  k_y} A_k \exp{[i(\bm{k} \cdot \bm{x} +\phi_{\bm{k}})]}$, where $\bm{k}
:=\{ k_x, k_y \}$ is the wavevector with modulus $k := |\bm{k}| =(2\pi/L)
m$, where $m$ is the dimensionless wavenumber, and $\phi_{\bm{k}}$ are
random phases uniformly distributed in $[0, 2\pi)$. The mode amplitudes
  are defined as $A_k := [1 + (k/k_0)^{15/3}]^{-1}$, peaking at $k_0 =
  (2\pi/L) m_0$ with $m_0 = 4$.  All modes with $m > 7$ are suppressed to
  ensure that the initial conditions are dominated by random, large-scale
  structures. The in-plane magnetic field components, $B_x$ and $B_y$,
  are then obtained from the vector potential via $\bm{B} = \nabla A_z
  \times \bm{\hat{z}}$. Finally, to probe a regime of strongly perturbed
  magnetic fields~\citep{Nattila2022}, the amplitude of fluctuations is
  set such that the root-mean-square of the in-plane magnetic field
  satisfies the condition $\langle B_\perp \rangle /B_0 \sim 1$, where
  $B_0$ is the mean magnetic field.


\bibliographystyle{abbrvnat}
\bibliography{aeireferences} 

@ARTICLE{Bacchini2019,
       author = {{Bacchini}, F. and {Ripperda}, B. and {Porth}, O. and {Sironi}, L.},
        title = "{Generalized, Energy-conserving Numerical Simulations of Particles in General Relativity. II. Test Particles in Electromagnetic Fields and GRMHD}",
      journal = {Astrophys. J., Supp.},
         year = "2019",
        month = feb,
       volume = {240},
       number = {2},
          eid = {40},
        pages = {40},
          doi = {10.3847/1538-4365/aafcb3},
archivePrefix = {arXiv},
       eprint = {1810.00842},
 primaryClass = {astro-ph.HE},
       adsurl = {https://ui.adsabs.harvard.edu/abs/2019ApJS..240...40B}}

@ARTICLE{bacchini2024,
       author = {{Bacchini}, Fabio and {Zhdankin}, Vladimir and {Gorbunov}, Evgeny A. and {Werner}, Gregory R. and {Arzamasskiy}, Lev and {Begelman}, Mitchell C. and {Uzdensky}, Dmitri A.},
        title = "{Collisionless Magnetorotational Turbulence in Pair Plasmas: Steady-State Dynamics, Particle Acceleration, and Radiative Cooling}",
      journal = {\prl},
         year = 2024,
        month = jul,
       volume = {133},
       number = {4},
          eid = {045202},
        pages = {045202},
          doi = {10.1103/PhysRevLett.133.045202},
archivePrefix = {arXiv},
       eprint = {2401.01399},
 primaryClass = {astro-ph.HE},
       adsurl = {https://ui.adsabs.harvard.edu/abs/2024PhRvL.133d5202B}
}

@ARTICLE{Ball2018,
   author = {{Ball}, D. and {{\"O}zel}, F. and {Psaltis}, D. and {Chan}, C.-K. and
	{Sironi}, L.},
    title = "{The Properties of Reconnection Current Sheets in GRMHD Simulations of Radiatively Inefficient Accretion Flows}",
  journal = {Astrophys. J.},
archivePrefix = "arXiv",
   eprint = {1705.06293},
 primaryClass = "astro-ph.HE",
     year = 2018,
    month = feb,
   volume = 853,
      eid = {184},
    pages = {184},
      doi = {10.3847/1538-4357/aaa42f},
   adsurl = {http://adsabs.harvard.edu/abs/2018ApJ...853..184B}}

@article{Bell1978,
	Adsurl = {http://adsabs.harvard.edu/abs/1978MNRAS.182..147B},
	Author = {{Bell}, A.~R.},
	Journal = {Mon. Not. R. Astron. Soc.},
	Month = jan,
	Pages = {147-156},
	Title = {{The acceleration of cosmic rays in shock fronts. I}},
	Volume = 182,
	Year = 1978}

@book{biskamp2003magnetohydrodynamic,
  title={Magnetohydrodynamic turbulence},
  author={Biskamp, Dieter},
  year={2003},
  publisher={Cambridge University Press}
}

@ARTICLE{Bruentti2007,
       author = {{Brunetti}, G. and {Lazarian}, A.},
        title = "{Compressible turbulence in galaxy clusters: physics and stochastic particle re-acceleration}",
      journal = {Mon. Not. R. Astron. Soc.},
         year = 2007,
        month = jun,
       volume = {378},
       number = {1},
        pages = {245-275},
          doi = {10.1111/j.1365-2966.2007.11771.x},
archivePrefix = {arXiv},
       eprint = {astro-ph/0703591},
 primaryClass = {astro-ph},
       adsurl = {https://ui.adsabs.harvard.edu/abs/2007MNRAS.378..245B}
}

@ARTICLE{Cerutti2013,
       author = {{Cerutti}, B. and {Werner}, G.~R. and {Uzdensky}, D.~A. and {Begelman}, M.~C.},
        title = "{Simulations of Particle Acceleration beyond the Classical Synchrotron Burnoff Limit in Magnetic Reconnection: An Explanation of the Crab Flares}",
      journal = {Astrophys. J.},
         year = 2013,
        month = jun,
       volume = {770},
       number = {2},
          eid = {147},
        pages = {147},
          doi = {10.1088/0004-637X/770/2/147},
archivePrefix = {arXiv},
       eprint = {1302.6247},
 primaryClass = {astro-ph.HE},
       adsurl = {https://ui.adsabs.harvard.edu/abs/2013ApJ...770..147C}
}

@ARTICLE{Chandran2000,
       author = {{Chandran}, Benjamin D.~G.},
        title = "{Scattering of Energetic Particles by Anisotropic Magnetohydrodynamic Turbulence with a Goldreich-Sridhar Power Spectrum}",
      journal = {Phys. Rev. Lett.},
         year = 2000,
        month = nov,
       volume = {85},
       number = {22},
        pages = {4656-4659},
          doi = {10.1103/PhysRevLett.85.4656},
archivePrefix = {arXiv},
       eprint = {astro-ph/0008498},
 primaryClass = {astro-ph},
       adsurl = {https://ui.adsabs.harvard.edu/abs/2000PhRvL..85.4656C}
}

@ARTICLE{Chen2018,
       author = {{Chen}, Liang},
        title = "{On the Jet Properties of {\ensuremath{\gamma}}-Ray-loud Active Galactic Nuclei}",
      journal = {Astrophys. J., Supp.},
         year = 2018,
        month = apr,
       volume = {235},
       number = {2},
          eid = {39},
        pages = {39},
          doi = {10.3847/1538-4365/aab8fb},
archivePrefix = {arXiv},
       eprint = {1803.05715},
 primaryClass = {astro-ph.HE},
       adsurl = {https://ui.adsabs.harvard.edu/abs/2018ApJS..235...39C}
}

@ARTICLE{Comisso2018,
       author = {{Comisso}, Luca and {Sironi}, Lorenzo},
        title = "{Particle Acceleration in Relativistic Plasma Turbulence}",
      journal = {Phys. Rev. Lett.},
         year = 2018,
        month = dec,
       volume = {121},
       number = {25},
          eid = {255101},
        pages = {255101},
          doi = {10.1103/PhysRevLett.121.255101},
archivePrefix = {arXiv},
       eprint = {1809.01168},
 primaryClass = {astro-ph.HE},
       adsurl = {https://ui.adsabs.harvard.edu/abs/2018PhRvL.121y5101C}}

@ARTICLE{Comisso2019,
       author = {{Comisso}, Luca and {Sironi}, Lorenzo},
        title = "{The Interplay of Magnetically Dominated Turbulence and Magnetic Reconnection in Producing Nonthermal Particles}",
      journal = {Astrophys. J.},
         year = 2019,
        month = dec,
       volume = {886},
       number = {2},
          eid = {122},
        pages = {122},
          doi = {10.3847/1538-4357/ab4c33},
archivePrefix = {arXiv},
       eprint = {1909.01420},
 primaryClass = {astro-ph.HE},
       adsurl = {https://ui.adsabs.harvard.edu/abs/2019ApJ...886..122C}
}

@ARTICLE{Dalena2014,
       author = {{Dalena}, S. and {Rappazzo}, A.~F. and {Dmitruk}, P. and {Greco}, A. and {Matthaeus}, W.~H.},
        title = "{Test-particle Acceleration in a Hierarchical Three-dimensional Turbulence Model}",
      journal = {Astrophys. J.},
         year = 2014,
        month = mar,
       volume = {783},
       number = {2},
          eid = {143},
        pages = {143},
          doi = {10.1088/0004-637X/783/2/143},
archivePrefix = {arXiv},
       eprint = {1402.3745},
 primaryClass = {astro-ph.SR},
       adsurl = {https://ui.adsabs.harvard.edu/abs/2014ApJ...783..143D}
}

@ARTICLE{Davelaar2019,
       author = {{Davelaar}, Jordy and {Olivares}, Hector and {Porth}, Oliver and
         {Bronzwaer}, Thomas and {Janssen}, Michael and {Roelofs}, Freek and
         {Mizuno}, Yosuke and {Fromm}, Christian M. and {Falcke}, Heino and
         {Rezzolla}, Luciano},
        title = "{Modeling non-thermal emission from the jet-launching region of M 87 with adaptive mesh refinement}",
      journal = {Astron. Astrophys.},
         year = "2019",
        month = dec,
       volume = {632},
          eid = {A2},
        pages = {A2},
          doi = {10.1051/0004-6361/201936150},
archivePrefix = {arXiv},
       eprint = {1906.10065},
 primaryClass = {astro-ph.HE},
       adsurl = {https://ui.adsabs.harvard.edu/abs/2019A&A...632A...2D}
}

@ARTICLE{Drake2009,
       author = {{Drake}, J.~F. and {Cassak}, P.~A. and {Shay}, M.~A. and {Swisdak}, M. and {Quataert}, E.},
        title = "{A Magnetic Reconnection Mechanism for Ion Acceleration and Abundance Enhancements in Impulsive Flares}",
      journal = {Astrophys. J. Lett.},
         year = 2009,
        month = jul,
       volume = {700},
       number = {1},
        pages = {L16-L20},
          doi = {10.1088/0004-637X/700/1/L16},
       adsurl = {https://ui.adsabs.harvard.edu/abs/2009ApJ...700L..16D}}

@ARTICLE{Fromm2022,
       author = {{Fromm}, C.~M. and {Mizuno}, Y. and {Younsi}, Z. and {Olivares}, H. and {Porth}, O. and {De Laurentis}, M. and {Falcke}, H. and {Kramer}, M. and {Rezzolla}, L.},
        title = "{Using space-VLBI to probe gravity around Sgr A$^{*}$}",
      journal = {Astron. Astrophys.},
         year = 2021,
        month = may,
       volume = {649},
          eid = {A116},
        pages = {A116},
          doi = {10.1051/0004-6361/201937335},
archivePrefix = {arXiv},
       eprint = {2101.08618},
 primaryClass = {astro-ph.HE},
       adsurl = {https://ui.adsabs.harvard.edu/abs/2021A&A...649A.116F}
}

@ARTICLE{Hillas1984,
       author = {{Hillas}, A.~M.},
        title = "{The Origin of Ultra-High-Energy Cosmic Rays}",
      journal = {Ann. Rev. Astron. \& Astrophys.},
         year = 1984,
        month = jan,
       volume = {22},
        pages = {425-444},
          doi = {10.1146/annurev.aa.22.090184.002233},
       adsurl = {https://ui.adsabs.harvard.edu/abs/1984ARA&A..22..425H}
}

@article{Imbrogno2024,
       author = {{Imbrogno}, Mario and {Meringolo}, Claudio and {Servidio}, Sergio and {Cruz-Osorio}, Alejandro and {Cerutti}, Beno{\^\i}t and {Pegoraro}, Francesco},
        title = "{Long-lived Equilibria in Kinetic Astrophysical Plasma Turbulence}",
      journal = {Astrophys. J. Lett.},
         year = 2024,
        month = sep,
       volume = {972},
       number = {1},
          eid = {L5},
        pages = {L5},
          doi = {10.3847/2041-8213/ad6b9d},
archivePrefix = {arXiv},
       eprint = {2408.02656},
 primaryClass = {physics.plasm-ph},
       adsurl = {https://ui.adsabs.harvard.edu/abs/2024ApJ...972L...5I}
}

@ARTICLE{Imbrogno2025,
       author = {{Imbrogno}, Mario and {Meringolo}, Claudio and {Cruz-Osorio}, Alejandro and {Rezzolla}, Luciano and {Cerutti}, Beno{\^\i}t and {Servidio}, Sergio},
        title = "{Turbulence and Magnetic Reconnection in Relativistic Multispecies Plasmas}",
      journal = {Astrophys. J. Lett.},
         year = 2025,
        month = sep,
       volume = {990},
       number = {2},
          eid = {L33},
        pages = {L33},
          doi = {10.3847/2041-8213/adfb4c},
archivePrefix = {arXiv},
       eprint = {2505.09700},
 primaryClass = {physics.plasm-ph},
       adsurl = {https://ui.adsabs.harvard.edu/abs/2025ApJ...990L..33I}
}

@article{balbus1998instability,
  title={Instability, turbulence, and enhanced transport in accretion disks},
  author={Balbus, Steven A and Hawley, John F},
  journal={Reviews of modern physics},
  volume={70},
  number={1},
  pages={1},
  year={1998},
  publisher={APS}
}

@ARTICLE{Isliker2017,
       author = {{Isliker}, Heinz and {Vlahos}, Loukas and {Constantinescu}, Dana},
        title = "{Fractional Transport in Strongly Turbulent Plasmas}",
      journal = {Phys. Rev. Lett.},
         year = 2017,
        month = jul,
       volume = {119},
       number = {4},
          eid = {045101},
        pages = {045101},
          doi = {10.1103/PhysRevLett.119.045101},
archivePrefix = {arXiv},
       eprint = {1707.01526},
 primaryClass = {physics.plasm-ph},
       adsurl = {https://ui.adsabs.harvard.edu/abs/2017PhRvL.119d5101I}
}

@ARTICLE{Kagan2015,
       author = {{Kagan}, D. and {Sironi}, L. and {Cerutti}, B. and {Giannios}, D.},
        title = "{Relativistic Magnetic Reconnection in Pair Plasmas and Its Astrophysical Applications}",
      journal = {Space Science Reviews},
         year = "2015",
        month = oct,
       volume = {191},
       number = {1-4},
        pages = {545-573},
          doi = {10.1007/s11214-014-0132-9},
archivePrefix = {arXiv},
       eprint = {1412.2451},
 primaryClass = {astro-ph.HE},
       adsurl = {https://ui.adsabs.harvard.edu/abs/2015SSRv..191..545K}}

@ARTICLE{Lemoine2025,
       author = {{Lemoine}, M. and {Bresci}, V. and {Gremillet}, L.},
        title = "{Particle acceleration up to the synchrotron burn-off limit in relativistic magnetized turbulence}",
      journal = {arXiv e-prints},
         year = 2025,
        month = sep,
          eid = {arXiv:2509.06437},
        pages = {arXiv:2509.06437},
          doi = {10.48550/arXiv.2509.06437},
archivePrefix = {arXiv},
       eprint = {2509.06437},
 primaryClass = {astro-ph.HE},
       adsurl = {https://ui.adsabs.harvard.edu/abs/2025arXiv250906437L}
}

@ARTICLE{Lynn2014,
       author = {{Lynn}, Jacob W. and {Quataert}, Eliot and {Chandran}, Benjamin D.~G. and {Parrish}, Ian J.},
        title = "{Acceleration of Relativistic Electrons by Magnetohydrodynamic Turbulence: Implications for Non-thermal Emission from Black Hole Accretion Disks}",
      journal = {Astrophys. J., Supp.},
         year = 2014,
        month = aug,
       volume = {791},
       number = {1},
          eid = {71},
        pages = {71},
          doi = {10.1088/0004-637X/791/1/71},
archivePrefix = {arXiv},
       eprint = {1403.3123},
 primaryClass = {astro-ph.HE},
       adsurl = {https://ui.adsabs.harvard.edu/abs/2014ApJ...791...71L}
}

@ARTICLE{Lyutikov2019,
       author = {{Lyutikov}, Maxim and {Temim}, Tea and {Komissarov}, Sergey and {Slane}, Patrick and {Sironi}, Lorenzo and {Comisso}, Luca},
        title = "{Interpreting Crab Nebula's synchrotron spectrum: two acceleration mechanisms}",
      journal = {Mon. Not. R. Astron. Soc.},
         year = 2019,
        month = oct,
       volume = {489},
       number = {2},
        pages = {2403-2416},
          doi = {10.1093/mnras/stz2023},
archivePrefix = {arXiv},
       eprint = {1811.01767},
 primaryClass = {astro-ph.HE},
       adsurl = {https://ui.adsabs.harvard.edu/abs/2019MNRAS.489.2403L}
}

@ARTICLE{Megale2025,
       author = {{Megale}, R. and {Cruz-Osorio}, A. and {Ficarra}, G. and {Imbrogno}, M. and {Meringolo}, C. and {Primavera}, L. and {Rezzolla}, L. and {Servidio}, S.},
        title = "{Measuring the properties of homogeneous turbulence in curved space-times}",
      journal = {Mon. Not. R. Astron. Soc.},
         year = 2025,
        month = dec,
       volume = {544},
       number = {2},
        pages = {2011-2023},
          doi = {10.1093/mnras/staf1873},
archivePrefix = {arXiv},
       eprint = {2509.04566},
 primaryClass = {gr-qc},
       adsurl = {https://ui.adsabs.harvard.edu/abs/2025MNRAS.544.2011M}
}

@ARTICLE{Meringolo2023,
       author = {{Meringolo}, Claudio and {Cruz-Osorio}, Alejandro and {Rezzolla}, Luciano and {Servidio}, Sergio},
        title = "{Microphysical Plasma Relations from Special-relativistic Turbulence}",
      journal = {Astrophys. J.},
         year = 2023,
        month = feb,
       volume = {944},
       number = {2},
          eid = {122},
        pages = {122},
          doi = {10.3847/1538-4357/acaefe},
archivePrefix = {arXiv},
       eprint = {2301.02669},
 primaryClass = {astro-ph.HE},
       adsurl = {https://ui.adsabs.harvard.edu/abs/2023ApJ...944..122M}
}

@ARTICLE{Meringolo2024,
       author = {{Meringolo}, C. and {Pucci}, F. and {Nistic{\'o}}, G. and {Pezzi}, O. and {Servidio}, S. and {Malara}, F.},
        title = "{Joint action of phase mixing and nonlinear effects in MHD waves propagating in coronal loops}",
      journal = {Astron. Astrophys.},
         year = 2024,
        month = aug,
       volume = {688},
          eid = {A12},
        pages = {A12},
          doi = {10.1051/0004-6361/202349094},
archivePrefix = {arXiv},
       eprint = {2312.15355},
 primaryClass = {astro-ph.SR},
       adsurl = {https://ui.adsabs.harvard.edu/abs/2024A&A...688A..12M}
}

@ARTICLE{Meringolo2025a,
       author = {{Meringolo}, Claudio and {Camilloni}, Filippo and {Rezzolla}, Luciano},
        title = "{Electromagnetic Energy Extraction from Kerr Black Holes: Ab Initio Calculations}",
      journal = {Astrophys. J. Lett.},
         year = 2025,
        month = oct,
       volume = {992},
       number = {1},
          eid = {L8},
        pages = {L8},
          doi = {10.3847/2041-8213/ae06a6},
archivePrefix = {arXiv},
       eprint = {2507.08942},
 primaryClass = {gr-qc},
       adsurl = {https://ui.adsabs.harvard.edu/abs/2025ApJ...992L...8M}
}

@ARTICLE{Meringolo2025b,
       author = {{Meringolo}, Claudio and {Rezzolla}, Luciano},
        title = "{FPIC: a new Particle-In-Cell code for stationary and axisymmetric black-hole spacetimes}",
      journal = {arXiv e-prints},
         year = 2026,
        month = feb,
          eid = {arXiv:2602.07452},
        pages = {arXiv:2602.07452},
          doi = {10.48550/arXiv.2602.07452},
archivePrefix = {arXiv},
       eprint = {2602.07452},
 primaryClass = {astro-ph.HE},
       adsurl = {https://ui.adsabs.harvard.edu/abs/2026arXiv260207452M}
}

@ARTICLE{Nathanail2021b,
       author = {{Nathanail}, Antonios and {Mpisketzis}, Vasilis and {Porth}, Oliver and {Fromm}, Christian M. and {Rezzolla}, Luciano},
        title = "{Magnetic reconnection and plasmoid formation in three-dimensional accretion flows around black holes}",
      journal = {Mon. Not. R. Astron. Soc.},
         year = 2022,
        month = jul,
       volume = {513},
       number = {3},
        pages = {4267-4277},
          doi = {10.1093/mnras/stac1118},
archivePrefix = {arXiv},
       eprint = {2111.03689},
 primaryClass = {astro-ph.HE},
       adsurl = {https://ui.adsabs.harvard.edu/abs/2022MNRAS.513.4267N}}

@ARTICLE{Nattila2022,
       author = {{N{\"a}ttil{\"a}}, Joonas and {Beloborodov}, Andrei M.},
        title = "{Heating of Magnetically Dominated Plasma by Alfv{\'e}n-Wave Turbulence}",
      journal = {Phys. Rev. Lett.},
         year = 2022,
        month = feb,
       volume = {128},
       number = {7},
          eid = {075101},
        pages = {075101},
          doi = {10.1103/PhysRevLett.128.075101},
archivePrefix = {arXiv},
       eprint = {2111.15578},
 primaryClass = {astro-ph.HE},
       adsurl = {https://ui.adsabs.harvard.edu/abs/2022PhRvL.128g5101N}
}

@ARTICLE{Parker1958,
       author = {{Parker}, E.~N. and {Tidman}, D.~A.},
        title = "{Suprathermal Particles}",
      journal = {Physical Review},
         year = 1958,
        month = sep,
       volume = {111},
       number = {5},
        pages = {1206-1211},
          doi = {10.1103/PhysRev.111.1206},
       adsurl = {https://ui.adsabs.harvard.edu/abs/1958PhRv..111.1206P}
}

@ARTICLE{Pecora2018,
       author = {{Pecora}, F. and {Servidio}, S. and {Greco}, A. and {Matthaeus}, W.~H. and {Burgess}, D. and {Haynes}, C.~T. and {Carbone}, V. and {Veltri}, P.},
        title = "{Ion diffusion and acceleration in plasma turbulence}",
      journal = {Journal of Plasma Physics},
         year = 2018,
        month = dec,
       volume = {84},
       number = {6},
          eid = {725840601},
        pages = {725840601},
          doi = {10.1017/S0022377818000995},
archivePrefix = {arXiv},
       eprint = {1803.09647},
 primaryClass = {physics.plasm-ph},
       adsurl = {https://ui.adsabs.harvard.edu/abs/2018JPlPh..84f7201P}}

@ARTICLE{Pezzi2024,
       author = {{Pezzi}, O. and {Trotta}, D. and {Benella}, S. and {Sorriso-Valvo}, L. and {Malara}, F. and {Pucci}, F. and {Meringolo}, C. and {Matthaeus}, W.~H. and {Servidio}, S.},
        title = "{Turbulence and particle energization in twisted flux ropes under solar-wind conditions}",
      journal = {Astron. Astrophys.},
         year = 2024,
        month = jun,
       volume = {686},
          eid = {A116},
        pages = {A116},
          doi = {10.1051/0004-6361/202348700},
archivePrefix = {arXiv},
       eprint = {2311.14428},
 primaryClass = {physics.plasm-ph},
       adsurl = {https://ui.adsabs.harvard.edu/abs/2024A&A...686A.116P}
}

@article{Radice2012b,
	Adsurl = {http://adsabs.harvard.edu/abs/2013ApJ...766L..10R},
	Archiveprefix = {arXiv},
	Author = {{Radice}, D. and {Rezzolla}, L.},
	Doi = {10.1088/2041-8205/766/1/L10},
	Eid = {L10},
	Eprint = {1209.2936},
	Journal = {Astrophys. J.},
	Month = mar,
	Pages = {L10},
	Primaryclass = {astro-ph.HE},
	Title = {{Universality and Intermittency in Relativistic Turbulent Flows of a Hot Plasma}},
	Volume = 766,
	Year = 2013}

@ARTICLE{Servidio2012,
       author = {{Servidio}, S. and {Valentini}, F. and {Califano}, F. and {Veltri}, P.},
        title = "{Local Kinetic Effects in Two-Dimensional Plasma Turbulence}",
      journal = {Phys. Rev. Lett.},
         year = 2012,
        month = jan,
       volume = {108},
       number = {4},
          eid = {045001},
        pages = {045001},
          doi = {10.1103/PhysRevLett.108.045001},
       adsurl = {https://ui.adsabs.harvard.edu/abs/2012PhRvL.108d5001S}}

@ARTICLE{Sironi2011,
       author = {{Sironi}, Lorenzo and {Spitkovsky}, Anatoly},
        title = "{Particle Acceleration in Relativistic Magnetized Collisionless Electron-Ion Shocks}",
      journal = {Astrophys. J.},
         year = 2011,
        month = jan,
       volume = {726},
       number = {2},
          eid = {75},
        pages = {75},
          doi = {10.1088/0004-637X/726/2/75},
archivePrefix = {arXiv},
       eprint = {1009.0024},
 primaryClass = {astro-ph.HE},
       adsurl = {https://ui.adsabs.harvard.edu/abs/2011ApJ...726...75S}
}

@article{Sironi2014,
   author = {{Sironi}, L. and {Spitkovsky}, A.},
    title = "{Relativistic Reconnection: An Efficient Source of Non-thermal Particles}",
  journal = {Astrophys. J.l},
archivePrefix = "arXiv",
   eprint = {1401.5471},
 primaryClass = "astro-ph.HE",
     year = 2014,
    month = mar,
   volume = 783,
      eid = {L21},
    pages = {L21},
      doi = {10.1088/2041-8205/783/1/L21},
   adsurl = {http://adsabs.harvard.edu/abs/2014ApJ...783L..21S}
}

@ARTICLE{Vega2024,
       author = {{Vega}, Cristian and {Boldyrev}, Stanislav and {Roytershteyn}, Vadim},
        title = "{Particle Acceleration in Relativistic Alfv{\'e}nic Turbulence}",
      journal = {Astrophys. J.},
         year = 2024,
        month = aug,
       volume = {971},
       number = {1},
          eid = {106},
        pages = {106},
          doi = {10.3847/1538-4357/ad5f8f},
archivePrefix = {arXiv},
       eprint = {2405.07891},
 primaryClass = {physics.plasm-ph},
       adsurl = {https://ui.adsabs.harvard.edu/abs/2024ApJ...971..106V}
}

@article{Vos2024,
       author = {{Vos}, Jesse and {Cerutti}, Beno{\^\i}t and {Mo{\'s}cibrodzka}, Monika and {Parfrey}, Kyle},
        title = "{Particle Acceleration in Collisionless Magnetically Arrested Disks}",
      journal = {Phys. Rev. Lett.},
         year = 2025,
        month = jul,
       volume = {135},
       number = {1},
          eid = {015201},
        pages = {015201},
          doi = {10.1103/5wwb-7t2n},
archivePrefix = {arXiv},
       eprint = {2410.19061},
 primaryClass = {astro-ph.HE},
       adsurl = {https://ui.adsabs.harvard.edu/abs/2025PhRvL.135a5201V}
}

@ARTICLE{Werner2016,
       author = {{Werner}, G.~R. and {Uzdensky}, D.~A. and {Cerutti}, B. and
         {Nalewajko}, K. and {Begelman}, M.~C.},
        title = "{The Extent of Power-law Energy Spectra in Collisionless Relativistic Magnetic Reconnection in Pair Plasmas}",
      journal = {Astrophys. J. Let.},
         year = "2016",
        month = jan,
       volume = {816},
       number = {1},
          eid = {L8},
        pages = {L8},
          doi = {10.3847/2041-8205/816/1/L8},
archivePrefix = {arXiv},
       eprint = {1409.8262},
 primaryClass = {astro-ph.HE},
       adsurl = {https://ui.adsabs.harvard.edu/abs/2016ApJ...816L...8W}}

@ARTICLE{Wong2020,
       author = {{Wong}, Kai and {Zhdankin}, Vladimir and {Uzdensky}, Dmitri A. and {Werner}, Gregory R. and {Begelman}, Mitchell C.},
        title = "{First-principles Demonstration of Diffusive-advective Particle Acceleration in Kinetic Simulations of Relativistic Plasma Turbulence}",
      journal = {Astrophys. J. Lett.},
         year = 2020,
        month = apr,
       volume = {893},
       number = {1},
          eid = {L7},
        pages = {L7},
          doi = {10.3847/2041-8213/ab8122},
archivePrefix = {arXiv},
       eprint = {1901.03439},
 primaryClass = {astro-ph.HE},
       adsurl = {https://ui.adsabs.harvard.edu/abs/2020ApJ...893L...7W}
}

@ARTICLE{Zank2015,
       author = {{Zank}, G.~P. and {Hunana}, P. and {Mostafavi}, P. and {Le Roux}, J.~A. and {Li}, Gang and {Webb}, G.~M. and {Khabarova}, O. and {Cummings}, A. and {Stone}, E. and {Decker}, R.},
        title = "{Diffusive Shock Acceleration and Reconnection Acceleration Processes}",
      journal = {Astrophys. J.},
         year = 2015,
        month = dec,
       volume = {814},
       number = {2},
          eid = {137},
        pages = {137},
          doi = {10.1088/0004-637X/814/2/137},
       adsurl = {https://ui.adsabs.harvard.edu/abs/2015ApJ...814..137Z}
}

@ARTICLE{Zhang2023,
       author = {{Zhang}, Hao and {Sironi}, Lorenzo and {Giannios}, Dimitrios and {Petropoulou}, Maria},
        title = "{The Origin of Power-law Spectra in Relativistic Magnetic Reconnection}",
      journal = {Astrophys. J. Lett.},
         year = 2023,
        month = oct,
       volume = {956},
       number = {2},
          eid = {L36},
        pages = {L36},
          doi = {10.3847/2041-8213/acfe7c},
archivePrefix = {arXiv},
       eprint = {2302.12269},
 primaryClass = {astro-ph.HE},
       adsurl = {https://ui.adsabs.harvard.edu/abs/2023ApJ...956L..36Z}
}

\end{document}